\documentclass[aps,prx,bibliography,superscriptaddress,twocolumn]{revtex4-2}
\usepackage[utf8]{inputenc}
\usepackage{amsmath}
\usepackage{physics}
\usepackage{mathtools}
\usepackage{amsfonts}
\usepackage{amssymb}
\usepackage{braket}
\usepackage{graphicx}
\usepackage{xr}
\usepackage{subfigure}
\usepackage{textcomp}
\usepackage{color}
\usepackage[dvipsnames]{xcolor}
\usepackage{mathrsfs}
\usepackage{natbib}
\usepackage{makecell}
\usepackage{soul}

\begin{document}
\title{Electric field induced Berry curvature dipole in quasi-one-dimensional Bi\textsubscript{4}I\textsubscript{4}}
\author{Amiya Mondal}
\email{amiyamondal@iisc.ac.in}
\affiliation{Solid State and Structural Chemistry Unit, Indian Institute of Science, Bangalore 560012, India}
\author{Awadhesh Narayan}
\email{awadhesh@iisc.ac.in}
\affiliation{Solid State and Structural Chemistry Unit, Indian Institute of Science, Bangalore 560012, India}
\date{\today}

\begin{abstract}
The nonlinear Hall effect in time-reversal symmetric materials offers a powerful probe into quantum geometry. Here, we investigate the electric-field-tunable nonlinear Hall response in few-layer $\text{Bi}_4\text{I}_4$ using comprehensive first-principles calculations across both its $\alpha$ and $\beta$ phases. Guided by symmetry analysis, we track the evolution of the Berry curvature dipole (BCD) tensor from the monolayer to the bilayer configuration under an out-of-plane electric field. While the monolayer features a highly rigid band structure and modest BCD tunability, the bilayer architecture exhibits substantial field-induced band modifications, including a progressive Rashba splitting and eventual gap closure in the $\beta$ phase. Crucially, this field-tunability allows substantial enhancement of the BCD magnitude relative to the monolayer counterpart. Our findings establish quasi-one-dimensional bismuth halogenides as a promising platform for engineering nonlinear Hall response.
\end{abstract}
\maketitle

\section{Introduction}

It is well established that the observation of the linear Hall effect necessitates the breaking of time-reversal symmetry~\cite{klitzing1980new}. Recently, however, an increasing body of research has demonstrated that nonlinear Hall characteristics can manifest in non-centrosymmetric materials, even when time-reversal symmetry remains strictly preserved~\cite{du2021nonlinear,ortix2021nonlinear,bandyopadhyay2024non}. This phenomenon was established on a rigorous theoretical foundation by Sodemann and Fu, who identified that a nonlinear Hall effect (NLHE) arises in such systems through the BCD mechanism~\cite{sodemann2015quantum}. The BCD is formally defined as the first momentum-space moment of the Berry curvature and serves to quantify its asymmetric distribution across the Brillouin zone.

Crystalline symmetries play a key role in determining the BCD and the resulting NLHE. In transition metal dichalcogenides (TMDs), the application of uniaxial strain reduces the point-group symmetry, thereby inducing a pronounced enhancement of the BCD~\cite{you2018berry,son2019strain}. A similar symmetry-lowering effect occurs in few-layer TMDs via varied stacking configurations, which has been shown to lead to comparable BCD enhancements~\cite{ma2019observation,kang2019nonlinear,huang2021room,joseph2021topological}. External perturbations offer additional avenues for dynamic control. For example, out-of-plane electric fields can effectively modulate the NLHE in low-dimensional materials~\cite{zhang2018electrically,bandyopadhyay2022electrically}, while pressure-induced topological phase transitions in Rashba materials such as BiTeI are predicted to generate giant BCD values~\cite{facio2018strongly}.

The BCD-driven NLHE represents a second-order transport phenomenon triggered by an in-plane electric field. The coupling between the applied field and the BCD manifests in a diverse array of physical consequences, ranging from giant magneto-optical responses and orbital valley magnetization~\cite{son2019strain} to the nonlinear Nernst~\cite{yu2019topological,zeng2019nonlinear} and thermal Hall effects~\cite{zeng2020fundamental}.

Significant concentration of Berry curvature typically occurs in regions of the Brillouin zone where energy bands exhibit near-degeneracy. In these narrow-gap regimes, the character of the Bloch states undergoes rapid variation in momentum space, making such systems ideal candidates for realizing a substantial BCD~\cite{kumar2021room,zeng2021nonlinear,matsyshyn2019nonlinear,singh2020engineering,roy2022non}. Consequently, materials characterized by small band gaps that are susceptible to external perturbations represent a highly promising avenue for exploring and engineering the BCD-induced NLHE.

As a tensor quantity, the BCD is strictly governed by crystalline symmetry constraints, with its non-vanishing components emerging only under specific broken symmetries. Consequently, low-symmetry systems represent the most promising avenue for observing BCD-induced NLHE. In this regard, quasi-one-dimensional materials such as $\text{Bi}_4\text{X}_4$ ($\text{X} = \text{Br, I}$) offer a unique and compelling platform for investigating the NLHE. Their quasi-one-dimensional characteristics originate from a chain-like monolayer architecture, where the intralayer binding energies are comparable to typical van der Waals interactions~\cite{zhou2014large}. These materials have been theoretically predicted and experimentally confirmed to host diverse topological phases across various stacking configurations~\cite{zhou2014large,zhou2015topological,peng2021observation,yang2022large,zhuang2021epitaxial}. For instance, monolayer $\alpha\text{-Bi}_4\text{Br}_4$ is a predicted quantum spin Hall insulator (QSHI)~\cite{zhou2014large,peng2021observation,zhang2017first}, while strain-induced topological edge states have been reported in monolayer $\beta\text{-Bi}_4\text{I}_4$~\cite{zhuang2021epitaxial}. 

Driven by advancements in synthesis and characterization techniques, recent experiments have further uncovered the rich topological landscapes of these systems. For example, recent scanning tunneling microscopy investigations coupled with first-principles calculations established the bulk $\alpha\text{-Bi}_4\text{I}_4$ crystal as a non-symmetry-indicated QSHI~\cite{yu2024observation}, characterized by non-zero spin Chern numbers~\cite{lin2024spin,deng2022twisted}. Furthermore, tracking the temperature evolution of the $(100)$ surface states of $\text{Bi}_4\text{I}_4$ via laser-based angle-resolved photoemission spectroscopy unveiled a thermal hysteresis of the surface gap across the $\alpha\text{-}\beta$ structural phase transition. This observation provided direct evidence of a temperature-induced topological phase transition from a weak topological insulator to a higher-order topological insulator phase~\cite{zhao2024topological}. As such, this class of materials, while well-studied for their topological properties, remains unexplored for their nonlinear Hall responses.

In the present work, we propose an electric field tunable nonlinear Hall response in few layer Bi$_4$I$_4$ systems. By employing first-principles calculations across both $\alpha$ and $\beta$ phases, we provide a comprehensive characterization of the nonlinear Hall response inherent to these quasi-one-dimensional systems. To complement our $ab-initio$ results, we perform symmetry analysis of both monolayer and bilayer configurations to delineate the structure of the BCD tensor. We examine the evolution of the BCD as a function of thickness and evaluate the influence of the external electric field on the modulation of the electronic structure. We find a substantial enhancement of the BCD in response to the applied field for the bilayer configuration in comparison to its monolayer counterpart. Overall, our work establishes quasi-one-dimensional bismuth halogenides as a tunable platform for exploring nonlinear Hall responses.

\section{Methodology}

Our first-principles calculations were conducted within the framework of density functional theory (DFT) using the {\sc quantum espresso} package~\cite{giannozzi2009quantum,giannozzi2017advanced}. Atomic structures were fully relaxed until the residual forces were reduced below $10^{-5}$ eV/\AA{}. We utilized projector-augmented wave pseudopotentials and the generalized gradient approximation as parameterized by Perdew, Burke, and Ernzerhof for the exchange-correlation functional~\cite{perdew1996generalized}. Given the presence of heavy elements in bismuth halide compounds, spin-orbit coupling (SOC) was included self-consistently across all calculation steps. An energy cutoff of 80 Ry was applied, with a $6 \times 6 \times 1$ $k$-mesh used for the Brillouin zone sampling. For the layered configurations, vacuum convergence was carefully evaluated for both $\alpha$ and $\beta$ phases in monolayer and bilayer forms. We used a minimum of 13 \AA{} optimized vacuum thickness to prevent spurious periodic interactions. To evaluate the Berry curvature and BCD properties, we constructed a tight-binding Hamiltonian based on maximally localized Wannier functions~\cite{marzari1997maximally} centered on the Bi and I $p$-orbitals, using the {\sc wannier90} code~\cite{mostofi2008wannier90}.

In systems preserving time-reversal symmetry, the nonlinear conductivity tensor, $\chi_{\alpha\beta\gamma}$, is governed by the momentum-space derivative of the Berry curvature, integrated over the occupied states~\cite{sodemann2015quantum},

\begin{equation}
\chi_{\alpha\beta\gamma} = \epsilon_{\alpha\delta\gamma} \frac{e^{3} \tau}{2(1+i\omega \tau)} \int [d\mathbf{k}] \, f_0 \left( \frac{\partial \Omega_{\delta}}{\partial k_{\beta}} \right),
\label{eq:chi}
\end{equation}

where $e$ denotes the electronic charge, $\tau$ is the scattering time, $\epsilon_{\alpha\delta\gamma}$ is the Levi-Civita symbol, and $f_0$ represents the equilibrium Fermi-Dirac distribution. The integration measure is defined as $[d\mathbf{k}] = d^{d}k/(2\pi)^d$ for a $d$-dimensional system. The BCD tensor is formally defined as~\cite{sodemann2015quantum}

\begin{equation}
D_{\alpha\beta} = \int [d\mathbf{k}] \, f_0 \left( \partial \Omega_{\beta} / \partial k_{\alpha} \right).
   \label{eq:bcd} 
\end{equation} 

Utilizing the Kubo formalism, the Berry curvature, $\Omega_{z}(\mathbf{k})$, is numerically evaluated as~\cite{xiao2010berry}

\begin{equation}
\Omega_{z}(\mathbf{k}) = 2i \sum_{i \neq j} \frac{\langle i | \partial H / \partial k_x | j \rangle \langle j | \partial H / \partial k_y | i \rangle}{(\varepsilon_i - \varepsilon_j)^2},
\label{eq:omega}
\end{equation}

where $\varepsilon_i$ and $\varepsilon_j$ represent the eigenenergies of the Hamiltonian $\hat{H}$ corresponding to the eigenstates $|i\rangle$ and $|j\rangle$, respectively. We note that in time-reversal invariant systems, the Berry curvature is an odd function of momentum, satisfying $T^{-1}\Omega_{z}(-\mathbf{k})T = -\Omega_{z}(\mathbf{k})$, where $T$ denotes the time-reversal operator. Conversely, the BCD tensor is even under time reversal, as $T^{\dagger}D_{\alpha\beta}(-\mathbf{k})T = D_{\alpha\beta}(\mathbf{k})$. The BCD calculations in our present study were performed using the {\sc wannier-berri} package, which provides an efficient implementation of the aforementioned methodology~\cite{tsirkin2021high}. 

\begin{figure}[b]
\centering
\includegraphics[width=0.40\textwidth]{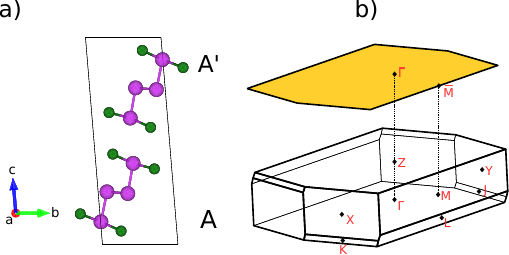}
\caption{Structure of bulk $\alpha$-Bi\textsubscript{4}I\textsubscript{4} showing (a) the side view of the unit cell and (b) illustration of the Brillouin zone, where the yellow shaded region is the two-dimensional projected Brillouin zone. The $\alpha$-Bi\textsubscript{4}I\textsubscript{4} monolayers are stacked along the $c$-axis with $AA'$ stacking. The purple and green spheres represent Bi and I atoms, respectively.}
\label{Fig_alpha-BiI-struc}
\end{figure}

\section{$\alpha$-B{\lowercase{i}}\textsubscript{4}I\textsubscript{4}}

\begin{figure*}[t]
\centering
\includegraphics[width=0.85\textwidth]{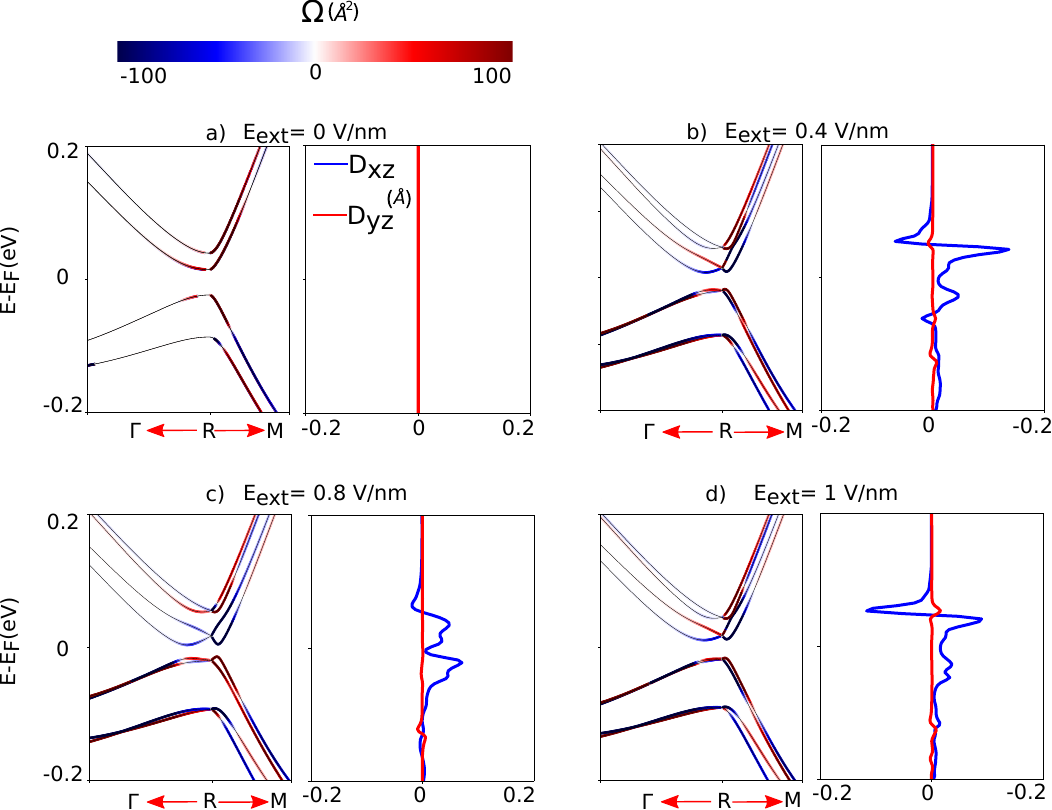}
\caption{Variation of band structure and BCD with applied electric field, E\textsubscript{ext}, for bilayer $\alpha$-Bi$_4$I$_4$. Vanishing BCD components can be seen in (a) zero applied field, while the BCD peaks emerging can be seen for (b) E\textsubscript{ext}=0.4 V/nm, (c) E\textsubscript{ext}=0.8 V/nm, and (d) E\textsubscript{ext}=1 V/nm. These BCD peak magnitudes increase with increasing E\textsubscript{ext}. The color gradient superimposed on the band structure represents the Berry curvature value.}
\label{Fig_bi-alpha-BiI}
\end{figure*}

In this section, we first present the structural properties of $\alpha$-Bi$_4$I$_4$ and the stacking arrangement of its layers. Subsequently, we explore the electronic structure and BCD response. Our discussion focuses primarily on the evolution of the electronic band structure and BCD components under an E\textsubscript{ext}. By applying a variable E\textsubscript{ext} ranging from $0$ to $1.0$ V/nm across both monolayer and bilayer configurations, we elucidate the mechanism of crystal symmetry breaking and its impact on the band structure and BCD magnitude.

\begin{figure}[t]
\centering
\includegraphics[width=0.48\textwidth]{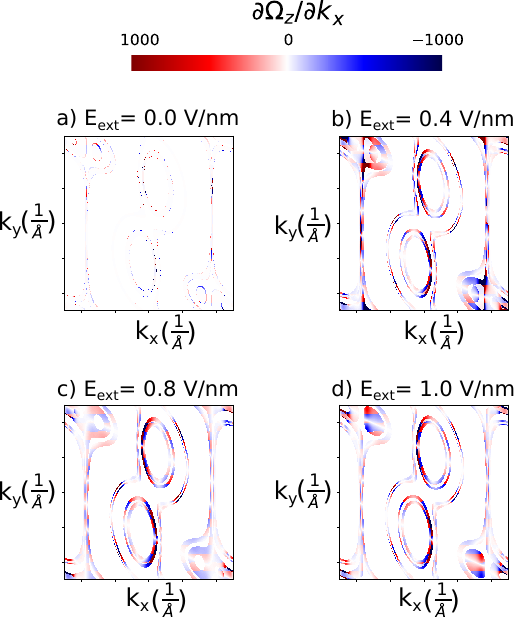}
\caption{Distribution of the Berry curvature derivative $\partial_{x}\Omega_{z}$ calculated on a constant energy contour (E-E\textsubscript{F}= 0.8 eV) for bilayer $\alpha$-Bi$_4$I$_4$ \textcolor{blue}. (a) At zero external field (E\textsubscript{ext}= 0.0 V/nm), the intensity is minimal. The application of an external field E\textsubscript{ext} $\neq$ 0 leads to a progressive amplification of the derivative density as shown in panels (b)–(d).}
\label{Fig_bi-alpha-dB}
\end{figure}

\subsection{Structural properties}

Bi$_4$I$_4$ typically stabilizes in the $\alpha$ phase, which crystallizes in the monoclinic space group $C_{2h}^3$ (C2/m)~\cite{liu2016weak}. The bulk unit cell is characterized by an $AA'$ stacking sequence of quasi-one-dimensional monolayers. The experimentally reported lattice parameters are $a = 14.213$~\AA, $b = 4.423$~\AA, $c = 19.939$~\AA, and $\beta = 92.91^\circ$~\cite{noguchi2019weak,huang2021room}. In this configuration, adjacent layers are coupled via weak van der Waals interactions. Within the bulk unit cell, the constituent chains in the bottom layers maintain a consistent arrangement and orientation. In contrast, the top layers are shifted by half a lattice vector along the crystallographic $b$-axis relative to the underlying layers, as depicted in Fig.~\ref{Fig_alpha-BiI-struc}(a). In the present study, we investigate both the monolayer and bilayer systems derived from this structural motif, focusing on the electronic properties and nonlinear Hall response, as we discuss in the following sections.

\subsection{Electronic structure and Berry curvature dipole}

\subsubsection{Bilayer results}

We begin by looking at the bilayer $\alpha$-Bi$_4$I$_4$ system, which comprises of two monolayers stacked along the crystallographic $c$-axis in an $AA'$ configuration. In this arrangement, the top and bottom layers maintain identical atomic orientations. However, the upper layer is translated by half a lattice vector along the $b$-axis relative to the lower layer, as illustrated in Fig.~\ref{Fig_alpha-BiI-struc}(a).

To determine the symmetry-allowed elements of the BCD tensor, we perform a formal symmetry analysis. Under a point-group symmetry operation represented by the matrix $S$, the BCD tensor $D$ transforms according to

\begin{equation}
D = \det(S) \, S D S^{-1},
\label{eq:bcd_symmetry}
\end{equation}

where $\det(S)$ accounts for the pseudotensor nature of the dipole~\cite{sodemann2015quantum}. In its pristine state, the bilayer $\alpha$-Bi$_4$I$_4$ possesses inversion symmetry, which enforces a vanishing BCD. However, as demonstrated below, the application of an out-of-plane E\textsubscript{ext} breaks this inversion symmetry, providing a mechanism to engineer a finite BCD. Based on symmetry constraints under the broken-symmetry configuration, we identify $D_{xz}$ and $D_{yz}$ as the only non-vanishing components of the BCD tensor.

In the absence of an external electric field (E\textsubscript{ext} = 0 V/nm), the bilayer exhibits a direct band gap of approximately 0.05 eV. This reduction in the gap magnitude relative to its monolayer counterpart is a direct consequence of enhanced inter-layer coupling. 

To investigate the tunability of the electronic structure, we apply an external electric field E\textsubscript{ext} along the out-of-plane direction, varying the magnitude from 0.2 to 1.0 V/nm. As illustrated in Fig.~\ref{Fig_bi-alpha-BiI}, the band structure reveals substantial Rashba-like splitting~\cite{bychkov1984properties} around the R point in both the valence and conduction bands. This splitting stems from the strong SOC inherent to the Bi atoms, combined with the field-induced breaking of inversion symmetry. Crucially, the magnitude of the Rashba splitting increases progressively with E\textsubscript{ext}, as evidenced by comparing Figs.~\ref{Fig_bi-alpha-BiI}(a) and (b). 

Examining the momentum-resolved Berry curvature along with the band structures in Fig.~\ref{Fig_bi-alpha-BiI}, the calculated Berry curvature exhibits prominent intensity localized near the band-crossing and near-degenerate regions. Away from these regions, the Berry curvature decays rapidly. This behavior is well-explained by Eq.~\eqref{eq:omega}, where the Berry curvature magnitude scales inversely with the square of the energy eigenvalue difference, maximizing its value near degenerate points. Interestingly, increasing E\textsubscript{ext} does not significantly alter the fundamental band gap magnitude. Instead, it alters the distribution and magnitude of the Berry curvature of the bands, with noticeable modifications appearing in the lower conduction bands [see Figs.~\ref{Fig_bi-alpha-BiI}(b) and (c)].

Significantly, the sensitivity of the BCD to the external field in the bilayer configuration deviates markedly from the monolayer behavior (discussed later). While the BCD components remain negligible at E\textsubscript{ext} = 0.2 V/nm, increasing the field strength up to E\textsubscript{ext} = 1.0 V/nm triggers a substantial enhancement of the BCD magnitudes, particularly of the $D_{xz}$ component. The BCD peaks are systematically aligned with the energy levels corresponding to the band-crossing points shown in Fig.~\ref{Fig_bi-alpha-BiI} within both the conduction and valence bands. Conversely, as the Fermi energy level approaches the band gap, the BCD magnitude falls sharply across all values of E\textsubscript{ext}. 

We anticipate that a further increase in E\textsubscript{ext} could continuously narrow the band gap, potentially driving the system toward a semi-metallic or metallic regime. To gain deeper microscopic insights, we evaluated the momentum-resolved derivative component $\partial_{x}\Omega_{z}$ at a constant energy of E-E\textsubscript{F}= 0.8 eV across varying E\textsubscript{ext}, given that $D_{xz}$ remains the dominant BCD contribution. This quantity represents the momentum-space gradient of the Berry curvature weighted by the distribution function, thereby defining the macroscopic BCD density. A detailed examination of Fig.~\ref{Fig_bi-alpha-dB} reveals a high concentration of $\partial_{x}\Omega_{z}$ intensity localized on the bands as the external field increases. This enhancement directly correlates with, and explains, our predicted field-induced increase of the BCD in this system.

\subsubsection{Monolayer results}

In its pristine state, monolayer $\alpha$-Bi$_4$I$_4$ is a narrow-gap insulator, exhibiting a direct band gap of approximately 0.07 eV upon the inclusion of SOC, see Fig.~S1(a) of the supplemental material~\cite{supplement}, which is notably larger than that of the bilayer configuration. As established in the preceding section, the broken inversion symmetry in this two-dimensional system is a prerequisite for achieving a finite BCD-induced NLHE. We note that the pristine monolayer structure shares the same essential symmetry operations as its bilayer counterpart.

As illustrated in Fig.~S1, the direct band gap at the high-symmetry R point remains remarkably robust under an applied electric field~\cite{supplement}. However, a detailed inspection of the band structure reveals substantial Rashba-like splitting~\cite{bychkov1984properties} near the R point in both the valence and conduction bands, resembling the behavior observed in the bilayer configuration.

In the absence of an external field (E\textsubscript{ext} = 0 V/nm), all BCD components vanish identically throughout the energy range, as shown in Fig.~S1(a). Upon the application of a finite E\textsubscript{ext} [Fig.~S1(b)], the $D_{xz}$ and $D_{yz}$ components become non-zero. Notably, in this monolayer configuration, E\textsubscript{ext} does not induce significant modulation of either the band gap or the absolute BCD magnitudes. Furthermore, these BCD values are far smaller than those observed in the bilayer case, highlighting a pronounced field-induced BCD response that is uniquely facilitated by the bilayer architecture. 

To further elucidate this behavior, we examine the momentum-space distribution of the Berry curvature derivative ($\partial_{a}\Omega_{b}$) on the Fermi surface across varying E\textsubscript{ext} (see Fig.~S2). We focus specifically on the $\partial_{x}\Omega_{z}$ component, as $D_{xz}$ represents the dominant BCD contribution relative to $D_{yz}$ (Fig.~S1). Due to the Rashba splitting in the electronic structure, the constant-energy contours manifest as concentric circles hosting a high intensity of $\partial_{x}\Omega_{z}$ compared to the other bands. Because these Rashba-split states reside immediately below and above the band gap, they generate small BCD peaks at their corresponding energy levels. 

We note that the contrasting field-responsive behaviors observed between the monolayer and bilayer systems can be attributed to the additional structural degrees of freedom inherent to the multi-layer configuration. As such, this enables a more effective field-induced band modulation.

\begin{figure}[t]
\centering
\includegraphics[width=0.48\textwidth]{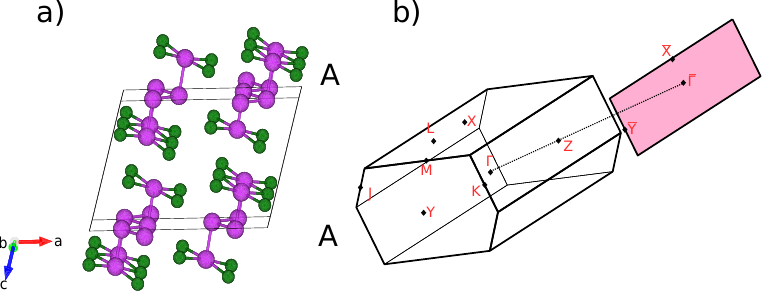}
\caption{Structure of bulk $\beta$-Bi\textsubscript{4}I\textsubscript{4} showing (a) the side view of the unit cell and (b) illustration of the Brillouin zone, where the pink shaded region is two-dimensional projected Brillouin zone. The $\beta$-Bi\textsubscript{4}I\textsubscript{4} monolayers are stacked along the $c$-axis with $AA$ stacking. The purple and green spheres depict Bi and I atoms, respectively.}
\label{Fig_beta-BiI-struc}
\end{figure}

\section{$\beta$-B{\lowercase{i}}\textsubscript{4}I\textsubscript{4}}

\begin{figure*}[t]
\centering
\includegraphics[width=0.85\textwidth]{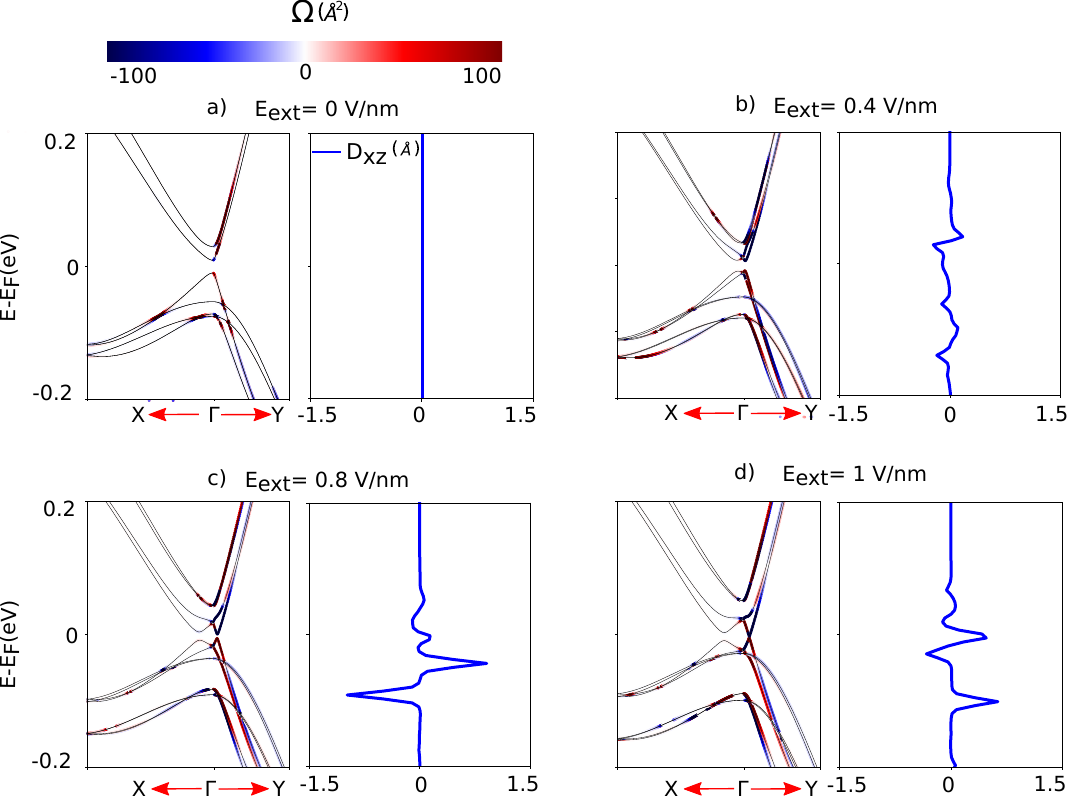}
\caption{Variation of the electronic band structure and BCD with the applied E\textsubscript{ext} for bilayer $\beta$-Bi$_4$I$_4$. (a) In the absence of E\textsubscript{ext}, the BCD vanishes due to inversion symmetry. (b)-(d) A finite BCD component, $D_{xz}$, emerges under a non-zero electric field. The color gradient superimposed on the band structure represents the calculated momentum-resolved Berry curvature.}
\label{Fig_bi-beta-BiI}
\end{figure*}

We next turn our attention to the $\beta$ phase. This section details the structural characteristics of $\beta$-Bi$_4$I$_4$, focusing on the unique stacking sequence of its constituent monolayers within the unit cell. Following this, we evaluate its electronic properties and BCD characteristics. We track the evolution of the electronic band structures and BCD components under an out-of-plane E\textsubscript{ext}. We investigate the variation of the BCD properties of this phase by systematically varying E\textsubscript{ext} for both monolayer and bilayer geometries.

\subsection{Structural properties}

The $\beta$ phase of Bi$_4$I$_4$ crystallizes in the same monoclinic space group as the $\alpha$ phase but adopts a distinct stacking configuration. Unlike the $AA'$ stacking of the $\alpha$ phase, the $\beta$ phase features an $AA$ stacking sequence of monolayers within the bulk unit cell, as depicted in Fig.~\ref{Fig_beta-BiI-struc}(a). The lattice parameters are $a = 14.386$~\AA, $b = 4.430$~\AA, $c = 10.493$~\AA, and $\beta = 107.9^\circ$~\cite{noguchi2019weak,huang2021room,autes2016novel}. The $\beta$ phase is metastable and has been observed to readily transform into the $\alpha$ phase upon cooling from growth temperature to ambient conditions.

\subsection{Electronic structure and Berry curvature dipole}

\subsubsection{Bilayer results}

As before, we begin by considering the bilayer structure. The bilayer architecture is constructed via an $AA$-type stacking sequence of monolayers along the crystallographic $c$-direction, as illustrated in Fig.~\ref{Fig_beta-BiI-struc}(a). In its pristine state, the crystal symmetry of bilayer $\beta$-Bi$_4$I$_4$ comprises a twofold rotation axis $C_{2y}$, a mirror plane $M_{xz}$, and inversion symmetry. However, upon the application of an E\textsubscript{ext} along the stacking direction, both the $C_{2y}$ rotation and inversion symmetries are explicitly broken. By applying the symmetry constraints defined in Eq.~\eqref{eq:bcd_symmetry}, we identify $D_{xz}$ as the sole non-vanishing BCD component. Specifically, the $D_{yz}$ element is strictly constrained to zero by the remaining mirror symmetry $M_{xz}$, a constraint consistent with the maximal symmetry requirements established for two-dimensional systems~\cite{sodemann2015quantum}. This behavior presents a stark contrast to the $\alpha$ phase, where symmetry considerations allow both independent BCD components to coexist. We note that this symmetry distinction could potentially be used to distinguish between the $\alpha$ and $\beta$ phases via nonlinear transport experiments.

We now examine the field-dependent modulation of this bilayer system. Similar to the $\alpha$ phase, the application of E\textsubscript{ext} induces a notable Rashba-like splitting that scales progressively with the field strength, as depicted in Figs.~\ref{Fig_bi-beta-BiI}(a) and (b). Concurrently, a significant reduction and eventual closure of the electronic band gap occur under increasing $E_{\text{ext}}$ (see Fig.~\ref{Fig_bi-beta-BiI}). 

\begin{figure}[t]
\centering
\includegraphics[width=0.48\textwidth]{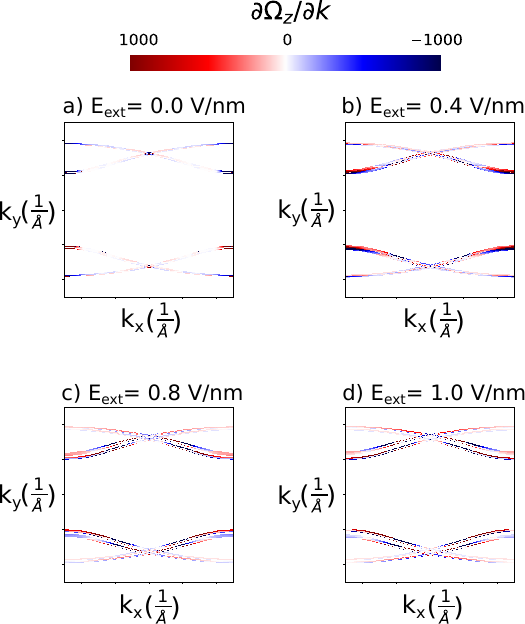}
\caption{Momentum-space distribution of the Berry curvature derivative $\partial_{x}\Omega_{z}$ calculated on a constant energy contour (E-E\textsubscript{F}= 0.8 eV) for bilayer $\beta$-Bi$_4$I$_4$. (a) At zero external field (E\textsubscript{ext}= 0.0 V/nm), the intensity is minimized, adhering to symmetry requirements. (b)–(d) With increasing E\textsubscript{ext}, a progressive amplification of the derivative density occurs alongside enhanced field-induced band splitting.}
\label{Fig_bi-beta-dB}
\end{figure}

To gain deeper microscopic insights into this behavior, we analyze the momentum-space distribution of the Berry curvature derivative ($\partial_{x}\Omega_{z}$) as a function of E\textsubscript{ext}. As shown in Fig.~\ref{Fig_bi-beta-dB}, the intensity of $\partial_{x}\Omega_{z}$ is significantly enhanced in the bilayer configuration relative to the monolayer $\beta$ phase (see supplement for a comparison to the monolayer~\cite{supplement}). A detailed inspection reveals that the high-intensity regions of $\partial_{x}\Omega_{z}$ observed in Fig.~\ref{Fig_bi-beta-dB}(b) correspond directly to the prominent BCD peak identified near the band-edge regions in Fig.~\ref{Fig_bi-beta-BiI}(b). In addition to the stark amplification of the $\partial_{x}\Omega_{z}$ intensity, the progressive splitting of the bands under stronger fields highlights the highly tunable nature of the underlying electronic structure. A comparison reveals that the high-intensity regions of the BCD distribution in Fig.\ref{Fig_bi-beta-dB} correspond directly to the prominent BCD peak seen in the band structure analysis of Fig.~\ref{Fig_bi-beta-BiI}.

Consequently, the macroscopic BCD response exhibits exceptional electric field sensitivity. While the magnitude of $D_{xz}$ remains relatively modest at E\textsubscript{ext}= 0.2 V/nm, a pronounced peak emerges as the field is increased to E\textsubscript{ext}= 1.0 V/nm. Remarkably, the maximum magnitude of $D_{xz}$ in this regime significantly exceeds those computed for the other two configurations investigated in this work, establishing the bilayer $\beta$ phase as the most responsive platform for field-induced nonlinear transport. In a nutshell, our findings demonstrate that bilayer $\beta$-Bi$_4$I$_4$ offers a highly promising venue where both the band gap and the BCD can be simultaneously and effectively modulated via E\textsubscript{ext}.

\subsubsection{Monolayer Results}

Prior to assessing the field-dependent BCD properties of monolayer $\beta$-Bi$_4$I$_4$, we briefly comment on its underlying topological character. Both DFT computations and experimental observations have robustly classified bulk $\beta$-Bi$_4$I$_4$ as a weak topological insulator~\cite{autes2016novel,liu2016weak,zhuang2021epitaxial}. As weak topological phases emerge from the weak vertical coupling of quantum spin Hall layers~\cite{hasan2010colloquium}, an isolated monolayer of $\beta$-Bi$_4$I$_4$ intrinsically behaves as a two-dimensional topological insulator. Mirroring the behavior of the $\alpha$ phase, the pristine monolayer $\beta$ phase exhibits a narrow direct band gap of approximately 0.045 eV. Additionally, the point-group symmetry operations of the isolated monolayer are identical to those of the bilayer configuration.

We next examine the impact of an out-of-plane E\textsubscript{ext} on the electronic bands and the induced BCD~\cite{supplement}. As depicted in Fig.~S3, the fundamental band gap demonstrates notable rigidity against the applied field. Crucially, the field-driven gap closure characteristic of the bilayer architecture is entirely missing in the monolayer limit, highlighting the vital role of inter-layer degrees of freedom in providing electrostatic tunability.

While the energy gap remains stable, the BCD components demonstrate a more discernible response to E\textsubscript{ext}. In particular, a prominent peak in the $D_{xz}$ component develops under a field of E\textsubscript{ext}= 1.0 V/nm [see Fig.~S3(d)]. To unravel the microscopic origin of this response, we map the momentum-resolved distribution of the Berry curvature derivative ($\partial_{x}\Omega_{z}$) across different field strengths. Mirroring the modest BCD amplitudes plotted in Fig.~S3, the corresponding momentum-space maps (Fig.~S4) show that the local $\partial_{x}\Omega_{z}$ hotspots remain fairly faint throughout the scanned range of E\textsubscript{ext}. This weak response stands in stark contrast to the bilayer $\beta$ phase, where the $\partial_{x}\Omega_{z}$ profile is substantially amplified due to more effective field-induced band modifications. 

\section{Estimation of nonlinear Hall currents}

Having established the field-induced BCD in our candidate systems, we now estimate the expected magnitude of the nonlinear Hall response. An oscillating external electric field $\vec{\mathcal{E}}(t) = \text{Re}[\vec{\mathcal{E}}_{0}e^{i\omega t}]$ induces a second-order electrical current density, $\vec{j}(t) = \text{Re}[\vec{J}^{(0)} + \vec{J}^{(2\omega)}e^{2i\omega t}]$, in systems characterized by a non-vanishing BCD~\cite{sodemann2015quantum}. This second-order response can be decomposed into two distinct components, namely, a stationary rectified DC current, given by $J^{(0)}_{i} = \chi_{ijk}\mathcal{E}^{*}_{j}\mathcal{E}_{k}$, and a second-harmonic AC component, expressed as $J^{(2\omega)}_{i} = \chi_{ijk}\mathcal{E}_{j}\mathcal{E}_{k}$. 

Under time-reversal symmetric conditions, the nonlinear conductivity tensor $\chi_{ijk}$ is determined by the momentum-space derivative of the Berry curvature over the occupied states, as defined by the BCD in Eq.~\eqref{eq:chi}. Here, the tensor indices $i, j, k$ span the two-dimensional spatial coordinates $x, y$. The nonlinear conductivity $\chi_{ijk}$ strictly respects the point-group symmetry of the crystal lattice and vanishes identically in the presence of inversion symmetry. Consequently, for crystals that lack inversion symmetry, a transverse Hall-like current can emerge as a second-order response to a longitudinal driving field $\mathcal{E}_x$. 

Using semiclassical Boltzmann transport theory within a single-band relaxation-time approximation~\cite{sodemann2015quantum}, the intraband contribution to the second-harmonic current density up to second order in the driving electric field is given by

\begin{equation}
\vec{j}^{(2\omega)} = \frac{e^3 \tau}{2\hbar^2(1 + i\omega\tau)} \hat{z} \times \vec{\mathcal{E}} (D \cdot \vec{\mathcal{E}}),
\label{eq:j2w}
\end{equation}

where $\tau$ represents the carrier relaxation time. The nonlinear current is directly proportional to the BCD. To provide a concrete quantitative analysis, we estimate the BCD-induced current density for $\beta$-Bi$_4$I$_4$, where both the monolayer and bilayer configurations possess a mirror plane $M_{xz}$. The maximal symmetry of a two-dimensional crystal that permits a non-zero BCD is a single mirror line~\cite{sodemann2015quantum}. The mirror plane $M_{xz}$ enforces the anti-symmetry condition $\Omega_z (k_{x}, k_{y}) = -\Omega_z (-k_{x}, k_{y})$ in momentum space. Substituting this constraint into Eq.~\eqref{eq:bcd} yields a non-vanishing $D_{xz}$ component, while $D_{yz}$ is strictly constrained to zero. We focus our estimation on the bilayer $\beta$-Bi$_4$I$_4$ configuration as it exhibits simultaneous external field tunability of both the band gap and the BCD magnitude.

Following Eq.~\eqref{eq:j2w}, when the driving harmonic field is aligned along the direction of the BCD vector ($x$-axis), a transverse nonlinear Hall current density is generated. For a representative peak value of $D_{xz} \approx 1$~\AA, as observed in the bilayer $\beta$-Bi$_4$I$_4$ system under an external field [Fig.~\ref{Fig_bi-beta-BiI}(c)], we adopt typical experimental parameters of $\tau = 10^{-12}$~s and an AC field amplitude of $\mathcal{E}_x = 100$~V/m. This yields a total estimated nonlinear AC response of $j = j^{(2\omega)} \approx   1849.05 \text{ A/m}^2 \times D_{xz} \sim 10^{-7}\text{ A/m}$. This measurable net current density underscores the viability of bilayer $\beta$-Bi$_4$I$_4$ as a promising platform for observing and manipulating nonlinear Hall transport.

\section{Summary and Outlook}

In summary, this work provides a comprehensive characterization of the field-tunable nonlinear Hall response in few-layer $\text{Bi}_4\text{I}_4$ systems across both their $\alpha$ and $\beta$ phases. By combining first-principles calculations with rigorous symmetry analyses of both monolayer and bilayer configurations, we have delineated the precise structure of the BCD tensor and established these quasi-one-dimensional bismuth halogenides as promising platforms for engineering nonlinear Hall transport. Their low crystal symmetry makes them attractive candidates for probing NLHE.

Our first-principles computations reveal a pronounced layer-dependent response to an out-of-plane electric field E\textsubscript{ext}. In the monolayer configurations, the electronic structure remains largely rigid, showing no substantial field-driven modulation of the band gap or Rashba splitting. In sharp contrast, the bilayer architectures afford an enhanced degree of electronic tunability. In these multi-layer systems, the electrostatic coupling facilitates robust external field control over the band gap, Rashba splitting, and a substantial enhancement of the BCD magnitude relative to the monolayer counterparts. Furthermore, we demonstrate that E\textsubscript{ext} serves as an efficient tool to break specific crystal symmetries, thereby engineering or amplifying the NLHE response. 

To unravel the microscopic mechanism driving this BCD behavior, we mapped the momentum-space distribution of the Berry curvature derivative density on constant-energy surfaces. This analysis revealed high-intensity hotspots that directly dictate the observed macroscopic dipole enhancement. Taken together, our results highlight bilayer $\alpha/\beta\text{-Bi}_4\text{I}_4$ as compelling architectures for electric-field-controlled nonlinear Hall transport. We hope our findings motivate future investigations into nonlinear Hall transport in quasi-one-dimensional van der Waals materials.

\section*{Acknowledgments}

We thank A. Reja, N.B. Joseph, N. Devaraj, and A. Bose for useful discussions. A.M. acknowledges Indian Institute of Science for a fellowship. A.N. thanks DST CRG grant (CRG/2023/000114) for support.

\bibliography{ref.bib}
\vspace{0.5cm}

\renewcommand{\theequation}{A\arabic{equation}}
\renewcommand{\thesection}{A\arabic{section}}
\setcounter{equation}{0}
\end{document}